\documentclass[twocolumn,showpacs,preprintnumbers,amsmath,amssymb]{revtex4}
\usepackage{graphicx}
\usepackage[latin1]{inputenc}
\usepackage{shortvrb}
\usepackage{epsfig}
\usepackage{amssymb}
\usepackage{mathrsfs}
\usepackage{amsbsy}
\usepackage{graphics}
\usepackage{epsf}
\usepackage{amsmath}
\usepackage{amssymb}
\usepackage[usenames]{color}
\usepackage{pstricks}

\usepackage{indentfirst}

\newcommand{\beq}{\begin{equation}}
\newcommand{\eeq}{\vspace{0cm} \end{equation}}
\newcommand{\beqq}{\setlength\arraycolsep{2pt}\begin{eqnarray}}
\newcommand{\eeqq}{\vspace{0cm} \end{eqnarray}}

\newcommand{\hsp}{\hspace{1cm}}

\begin{document}

\title{Can Dark Matter Decay in Dark Energy?}

\author{S. H. Pereira} \email{spereira@astro.iag.usp.br}

\author{J. F. Jesus} \email{jfernando@astro.iag.usp.br}

\affiliation{Universidade de S\~ao Paulo -- Instituto de Astronomia,
  Geof\'\i sica e Ci\^encias Atmosf\'ericas \\
Rua do Mat\~ao, 1226 -- 05508-090 Cidade Universit\'aria, S\~ao Paulo, SP,
Brazil}

\bigskip
\begin{abstract}
We analyze the interaction between Dark Energy and Dark Matter from a thermodynamical perspective. By assuming they have different temperatures, we study the possibility of occurring a decay from Dark Matter into Dark Energy, characterized by a negative parameter $Q$. We find that, if at least one of the fluids has non vanishing chemical potential, for instance $\mu_x<0$ and $\mu_{dm}=0$ or $\mu_x=0$ and $\mu_{dm}>0$, the decay is possible, where $\mu_x$ and $\mu_{dm}$ are the chemical potentials of Dark Energy and Dark Matter, respectively. Using recent cosmological data, we find that, for a fairly simple interaction, the Dark Matter decay is favored with a probability of $\sim 93\%$ over the Dark Energy decay. This result comes from a likelihood analysis where only
background evolution has been considered.
\end{abstract}

\maketitle

\section{Introduction}

There is no doubt that two of the main challenges of the present cosmology concerns the understanding of the nature of the Dark Energy (DE) and Dark Matter (DM). The former is responsible for the present accelerated expansion of the universe, as indicated by type Ia Supernovae observations \cite{SN,ast05}, and represents about 70\% of its material content. The latter one acts exactly like the ordinary matter, but does not interact with it, except gravitationally, corresponding to about 25\% of the material content of the Universe (see \cite{DMrev} for a review). In a series of recent papers the possibility of a coupling between DM and DE has been considered, and we have three possibilities: DM decaying to DE \cite{das,Feng,DM}, DE decaying to DM \cite{DE1,DE2,DE3,DE4,abwa,pavon1,JesusEtAl08,CarvalhoEtAl92,wm,Jesus06,alclim05} or interaction in both directions \cite{guo,Ioav}. Interaction among these completely different fluids has some important consequences, like addressing the coincidence problem, for instance. The coincidence problem can be solved or alleviated on these models by assuming that the DE decays into DM, thus diminishing the difference between the densities of the two components through the evolution of the Universe. Thus, if one finds that it is the DM that decays into DE, obviously these models can not be used to alleviate the coincidence problem, but, on the other hand, it is the DM that dominates in the past, and as the decay occurs, the dark energy starts to dominate, as indicated by observational data. Thus, the possibility of interaction between DE and DM can be investigated and the final answer shall be given by the cosmological observations.

Although the observational constraints indicate the possibility that the decay can be in both directions, the last mechanism, namely the decay of DE into DM, has been only recently studied \cite{abwa,pavon1} from a thermodynamical point of view and it seems to be strongly favored by the second law, provided that the chemical potential of both components vanish. In this paper we show that, if the chemical potential of at least one of the fluids is not zero, the decay can occur from the DM to DE, with no violation of the second law. The inclusion of a non null chemical potential to the dark energy fluid is not a merely theoretical tool. As showed in some recent papers \cite{limasaulo}, a negative chemical potential is necessary in order to turn the phantom Dark Energy a physically real hypothesis from the thermodynamical point of view.

The structure of the paper is as follows. Section 2 reviews the basic theory of two interacting fluids with different temperatures. In section 3 we apply this method to the case of Dark Matter and Dark Energy interaction. In section 4, we discuss briefly the Le Ch\^atelier-Braun principle, which is related to the thermodynamical equilibrium of two fluids. In section 5 we analyze a simple interacting DE-DM model, and finally we conclude on section 6.

\section{Thermodynamics of two interacting fluids}

In the following we review the theory of two interacting fluids, as developed by Zimdahl \cite{zim}. The energy-momentum tensor of two perfect fluids (denoted by 1 and 2) is
\beq
T^{ik}=T^{ik}_1+T^{ik}_2\,,\label{eq1}
\eeq
with
\beq
T^{ik}_A=(p_A+\rho_A)u^i u^k + p_A g^{ik}\,,\quad (c\equiv 1)\label{eq2}
\eeq
where $\rho_A$ is the energy density and $p_A$ the equilibrium pressure of the species $A=1,2$. The 4-velocity $u^i$ is assumed to be the same for both fluids. The particle flow vector $N^i_A$ is defined as
\beq
N^i_A = n_A u^i\,,\label{eq3}
\eeq
where $n_A$ is the particle number density. The balance equation for the particle number assumes the form
\beq
N^i_{A;i}=\dot{n}_A+\Theta n_A = n_A\Gamma_A\,,\label{eq4}
\eeq
where $\Gamma_A$ is the rate of change of the number of particles $A$ and $\Theta\equiv u^i_{;i}$ is the fluid expansion rate. Particle production is characterized by $\Gamma_A >0$, particle decay by $\Gamma_A< 0$ and for $\Gamma_A=0$ we have separated particle number conservation.

Considering interaction between the fluids we have that the total energy-momentum tensor is conserved, but not each system separately. For a general source term $t^i_A$ in the energy-momentum tensor, we have
\beq
T^{ik}_{A;k}=t^i_A\,,\label{eq5}
\eeq
implying in the conservation equations
\beq
\dot{\rho}_A+\Theta(\rho_A+p_A)=u_a t^a_A\,,\label{eq6}
\eeq
\beq
(\rho_A+p_A)\dot{u}^a+p_{A,k}h^{ak}=-h^a_it^i_A\,,\label{eq7}
\eeq
where $h^{ij}=g^{ij}+u^iu^j$. Writing the entropy per particle as $s_A$ for each species, the Gibbs equation must be satisfied
\beq
T_A ds_A=d\frac{\rho_A}{n_A}+p_Ad\frac{1}{n_A}\,,\label{eq8}
\eeq
so that the entropy per particle variation rate is
\beq
\dot{s}_A=\frac{u_a t^a_A}{n_A T_A}-\frac{(\rho_A+p_A)}{n_A T_A}\Gamma_A\,.\label{eq9}
\eeq


The entropy flow vector $S^a_A$ is defined by
\beq
S^a_A=n_As_Au^a\,,\label{eq14}
\eeq
and we have
\begin{eqnarray}
S^a_{A;a}&=&n_As_A\Gamma_A+n_A\dot{s}_A \nonumber \\
 &=&\bigg(s_A-\frac{\rho_A+p_A}{n_A T_A}\bigg)n_A \Gamma_A+\frac{u_at^a_A}{T_A}\,\label{eq15}
\end{eqnarray}
where we have used relation (\ref{eq9}).

The condition of energy-momentum conservation, $T^{ik}_{;k}=0$, for the system as a whole implies $t_1^a=-t_2^a$, but there is no corresponding condition for the particle number as a whole, because the total number of particles can not be conserved. We have for the total particle number density
\beq
n=n_1+n_2\,\label{eq18}
\eeq
and
\beq
\dot{n}+\Theta n= n \Gamma\,,\label{eq19}
\eeq
where
\beq
n\Gamma = n_1\Gamma_1 + n_2\Gamma_2\,,\label{eq20}
\eeq
and $\Gamma$ is the rate by which the total particle number changes.

The entropy per particle is given by
\beq
s_A={\rho_A+p_A\over n_A T_A}-{\mu_A\over T_A}\,,\label{eq21}
\eeq
where $\mu_A$ is the chemical potential of the species $A$. Substituting in Eq. (\ref{eq15}) yields
\beq
S^a_{A;a}=-{\mu_A\over T_A}n_A \Gamma_A+{u_at^a_A\over T_A}\,.\label{eq22}
\eeq
The total entropy production density is
\begin{eqnarray}
S^a_{;a}&=&S^a_{1;a}+S^a_{2;a} \nonumber\\
&=&-\bigg({\mu_1\over T_1}n_1 \Gamma_1 +{\mu_2\over T_2}n_2 \Gamma_2 \bigg)+\bigg({u_at^a_1\over T_1}+{u_at^a_2\over T_2}\bigg)\,.\nonumber\\\label{eq24}
\end{eqnarray}

The equilibrium condition $S^a_{;a}=0$ requires $\mu_1=\mu_2$, $T_1=T_2$ and $t_1^a = -t_2^a$.

\section{Interaction between Dark Matter and Dark Energy}

Now let us apply the aforementioned results to the specific case of DM and DE interaction in the context of a FRW expansion. The Dark Energy density is represented by $\rho_x$ and is supposed to satisfy the equation of state (EoS)
\beq
p_x=\omega \rho_x\,.\label{eqst}
\eeq

The Dark Matter energy density is represented by $\rho_{dm}$ and its EoS can be approximately written as \cite{degroot}
\begin{equation}
\rho_{dm} = n_{dm}M + \frac{3}{2}n_{dm}T_{dm}\,
,\quad p_{dm} = n_{dm} \, T_{dm}\,
\end{equation}
where we have used $k_{B} = 1$. In the limit $T_{dm}\ll M$ these reduce to $\rho_{dm}\simeq n_{dm} M$ and $p_{dm}\simeq 0$. Defining the interaction term $u_at^a_{dm}\equiv Q$, the energy density conservation (\ref{eq6}) of each component turns
\begin{eqnarray}
\label{conerv2b}
\dot{\rho}_{dm}+3H \rho_{dm} &=& Q \, , \\
\dot{\rho}_{x}+3H(1+\omega)\rho_{x}&=& -Q \,.  \label{density}
\end{eqnarray}

For $Q>0$ we have the DE decaying into DM and for $Q<0$ is the DM that decays
into DE. The cosmological tests indicate as many positive values \cite{DE4} as negative values \cite{Feng} for $Q$.

Now let us analyze the process from a thermodynamical point of view. The total entropy production (\ref{eq24}) reads
\beq
S^a_{;a}=-\bigg({\mu_{dm}\over T_{dm}}n_{dm} \Gamma_{dm} +{\mu_x\over T_x}n_x \Gamma_x \bigg)+\bigg({1\over T_{dm}}-{1\over T_x}\bigg)Q\,.\label{entropy}
\eeq

Considering $\omega$ constant along the evolution, the temperature evolution for each component in terms of the scale factor is
\beq
T_{dm}\propto a^{-2}\,\,\,;\hsp T_x \propto a^{-3\omega}\,,
\eeq
thus, one expects that the present temperature of the Dark Energy  is much greater than the temperature of the Dark Matter ($T_x\gg T_{dm}$). Setting the chemical potentials as null ($\mu_{dm}=\mu_x=0$), we see that the positivity of the entropy production ($S^a_{;a}\geq 0$) implies $Q>0$, as already discussed in \cite{pavon1}.

But now let us consider that the chemical potentials are not null. For this case the condition of positivity of the entropy production becomes
\beq
Q\geq\bigg({\mu_x\over T_x}n_x\Gamma_x + {\mu_{dm}\over T_{dm}}n_{dm}\Gamma_{dm}\bigg){1\over (1/T_{dm} - 1/T_x)}\,.\label{cond}
\eeq
It is clear that negative values of $Q$ are also possible, depending only on the signs of the chemical potentials. For example, if one of the components has a negative chemical potential while the other is null, we can have negative values for $Q$.  Recently, the thermodynamical and statistical properties of conserved
Dark Energy fluids were reexamined by considering a
non-zero chemical potential ($\mu_x \neq 0$). It was found
that the entropy condition, $S^a_{;a} \geq 0$, implies that the possible
values of $\omega$ are heavily dependent on the value, as well as
on the sign of  the chemical potential \cite{limasaulo}. For $\mu_x >0$, the $\omega$-parameter would
be strictly greater than $-1$ (vacuum is forbidden) while for $\mu_x <
0$ not only the vacuum but even a phantom-like behavior ($\omega
<-1$) is allowed. In any case, the ratio between the chemical
potential and temperature remains constant for a conserved DE fluid, that is,
$\mu_x/T_x=\mu_{0x}/T_{0x}$, where $\mu_{0x}$, $T_{0x}$ are the present day values
of the chemical potential and temperature, respectively.

With basis on this discussion, taking for instance $\mu_{dm}=0$ and $\mu_x=-\alpha T_x$, with $\alpha$ a positive constant, the condition (\ref{cond}) turns
\beq
Q \gtrsim - \alpha n_x\Gamma_x T_{dm}\,.\label{cond2}
\eeq
This opens the possibility of negative values of $Q$, so that the decay from DM to DE is thermodynamically consistent. Observe that, if DE fluid is being created, the quantity $n_x\Gamma_x > 0$, according to (\ref{eq4}). It is also interesting to note that this relation depends on the temperature of the Dark Matter, which is expected to be presently a very tiny value.

As one may see, the expression (\ref{cond}) is symmetric on the indexes $x$ and $dm$ . Thus the same analysis is valid for the case $\mu_{dm}\neq 0$ and $\mu_x=0$. Indeed, a more general condition necessary to have negative values to $Q$, according to (\ref{cond}) is
\beq
{\mu_x\over T_x}n_x\Gamma_x <- {\mu_{dm}\over T_{dm}}n_{dm}\Gamma_{dm}\,.\label{condgeral}
\eeq
Remembering that $n_x\Gamma_x > 0$ and $n_{dm}\Gamma_{dm} < 0$ when DM is decaying into DE, for $\mu_x=0$ we must have $\mu_{dm}>0$.

Finally, a negative value of $Q$ is also allowed if we require that the entropy per particle must be conserved for each fluid separately. According to (\ref{eq9}) we have
\beq
Q=(\rho_{dm} + p_{dm})\Gamma_{dm}\simeq M n_{dm}\Gamma_{dm}\,,
\eeq
\beq
Q=-(\rho_{x} + p_{x})\Gamma_{x}\,.
\eeq
and the condition (\ref{condgeral}) turns
\beq
{\mu_x n_x\over T_x (1+\omega)\rho_x} < {\mu_{dm}\over M T_{dm}}\,.
\eeq

It is interesting to note that, in this case, the condition of DM decay does not depend on the decay rates, depending only of thermodynamical properties of each fluid.

\section{Le Ch\^atelier-Braun principle}

It is also important to understand the aforementioned process in the light of the Le Ch\^atelier-Braun principle \cite{callen}, as recently discussed by Pav\'on and Wang \cite{pavon1} for the case of DE decaying into DM ($Q>0$), both with null chemical potentials. The principle says that when a system is perturbed out of its equilibrium state it reacts in such manner to restore the equilibrium state or to achieve a new one. In the approach of null chemical potential discussed in \cite{pavon1}, today the answer of the system to the equilibrium loss is a continuous transfer of energy from DE to DM, corresponding to a flux $f$ from the former to the last one. This happens because if $Q>0$, $T_x$ will increase more slowly as the Universe expands than in the absence of interaction and correspondingly $T_{dm}$ will also decrease more slowly. This implies that the temperature difference between the fluids would decrease and eventually they will reach equilibrium. Indeed the equilibrium is never achieved because the expansion of the Universe acts as an external agent in the opposite direction.

Now let us see how this mechanism acts when the chemical potential is not null. Taking firstly the case with $\mu_x<0$ and $\mu_{dm}=0$, we see that the negative chemical potential of the DE acts as a {\it potential well}, ``attracting''  or ``sucking'' the DM particles to the DE sector. This represents a flux $F$ from DM to DE, in analogy to the previous case, but in opposite direction. Thus, if $|F|>|f|$, the net flux is in the DE direction, representing the decay of Dark Matter into Dark Energy, without violating the Le Ch\^atelier-Braun principle. The chemical potential acts as an ``attractor'' to the DM particles, or as an external agent driving the decay and implying in $Q<0$. The same happens in the case $\mu_x=0$ and $\mu_{dm}>0$. Now is the Dark Matter chemical potential that acts as a {\it potential hill}, ``repelling'' the DM particles to the DE sector, favoring the decay in the direction of DE, and again implying $Q<0$. Note that in this discussion we have not imposed $Q<0$. It follows from the action of the chemical potential, driving the direction in which the decay must occur.

\section{Application: $Q\propto H\rho_x$}
An interesting model of dark energy interacting with dark matter which has been recently proposed parametrizes the effect of the interaction on the dark matter evolution law \cite{JesusEtAl08}, with $Q\propto H\rho_{dm}$. However, although well succeeded in the background, this model is plagued with instabilities on the perturbations of the dark energy component \cite{Valiviita08}. As already shown on \cite{Abdalla08}, though, these instabilities are not a fairly general feature of these interacting models. They have shown that, in fact, models with $Q\propto H\rho_x$ can be free of such instabilities. We rely on the conclusions of \cite{Abdalla08}, which, although being given for the DE decay, can be shown to be valid for the DM decay, too, at least for this specific interacting model. 

By assuming that baryons are conserved, the energy conservation law for the two interacting components
($u_{\alpha}{T^{\alpha\beta}}_{{;}\beta}=0$) is given by Eqs. (\ref{conerv2b}) and (\ref{density}). At this point, we should mention that, although in this model the baryon contribution is not neglected, as usual on the literature, it shall not change the thermodynamical analysis discussed above, as baryons are separately conserved, thus it has $\Gamma_b\equiv0$, so it does not contribute to the inequality (\ref{cond}). Thus, the inclusion of baryons still allows for the decay of DM into DE, provided one of them has non-vanishing chemical potential. On another hand, the inclusion of baryons shall not spoil the stability of the perturbations, as the critical point for it is the interaction, of which baryons do not play a role.

Thus, we shall work with the following interaction term $Q$:
\begin{equation}
 Q=3\varepsilon H\rho_{x}\,.
\label{Q}
\end{equation}
where we have introduced the factor $3$ for convenience. From this equation we can see that the sign of $\varepsilon$ determines the sign of $Q$, at least while $\rho_x>0$. This model has already been studied on \cite{PavZim05,pavon1}.

With the interaction term (\ref{Q}) it is possible to integrate Eq. (\ref{density}), for a constant $\omega$, and show that the energy density of the DM and DE components are given by
\begin{eqnarray}
\rho_{dm} &=& \rho_{dm0}a^{-3}+ \frac{\varepsilon\rho_{x0}a^{-3}[1-a^{-3(\varepsilon+\omega)}]}{\varepsilon+\omega}\\
\rho_x &=& \rho_{x0}a^{-3(1+\varepsilon+\omega)}
\end{eqnarray}
where we can see that in the limit of vanishing $\varepsilon$, we recover the standard, non-interacting, behaviors. As baryons are conserved, we have $\rho_b=\rho_{b0}(1+z)^3$, so the total pressureless matter density $\rho_m$ $(=\rho_b+\rho_{dm})$ goes like:
\begin{equation}
\rho_m =\rho_{m0}a^{-3}+ \frac{\varepsilon\rho_{x0}a^{-3}[1-a^{-3(\varepsilon+\omega)}]}{\varepsilon+\omega}
\end{equation}

We choose to work on a spatially flat Universe, which is predicted by inflation and is in agreement with the (Wilkinson Microwave Anisotropy Probe) WMAP observations \cite{CMB,CMB2,Komatsu08}. Thus, neglecting the radiation contribution, the Friedmann equation for this modified XCDM cosmology can be written as:
\begin{eqnarray}
\label{friedmann} \left(\frac{{H}}{H_0}\right)^2 &=& \left[\Omega_m+\frac{\varepsilon\Omega_x}{\varepsilon+\omega}\right](1 + z)^3 +\nonumber\\
&+&\frac{\omega\Omega_x}{\omega + \varepsilon}(1 + z)^{3(1 + \varepsilon + \omega)}
\end{eqnarray}
where $H$, $\Omega_b$ and $\Omega_{m}$ $(=1-\Omega_x)$ are, respectively, the Hubble parameter, the baryons and matter present density parameters and we have dropped the subscript `0' for convenience. 

This model differs from the model of \cite{PavZim05,pavon1}, as they neglect the baryon contribution. Our aim here is to use the most recent observations and the inclusion of baryons in order to put stringent constraints on this model.

We fix $\Omega_b=0.0437$ from the WMAP 5-year observations \cite{Komatsu08}, which is also in agreement with the nucleosynthesis predictions, so we have four free parameters on our model, namely, $h$, $\Omega_m$ ($=\Omega_{dm}+\Omega_b$), $\omega$ and $\varepsilon$ (neglecting the radiation contribution, which is of order $\Omega_\gamma h^2 \sim 10^{-5}$ \cite{CMB2}). The dark energy density parameter $\Omega_x$ is not a free parameter as we choose to work on a spatially flat Universe. We shall combine various cosmological observations in order to constrain these parameters.

\begin{figure*}
\centerline{\epsfig{file=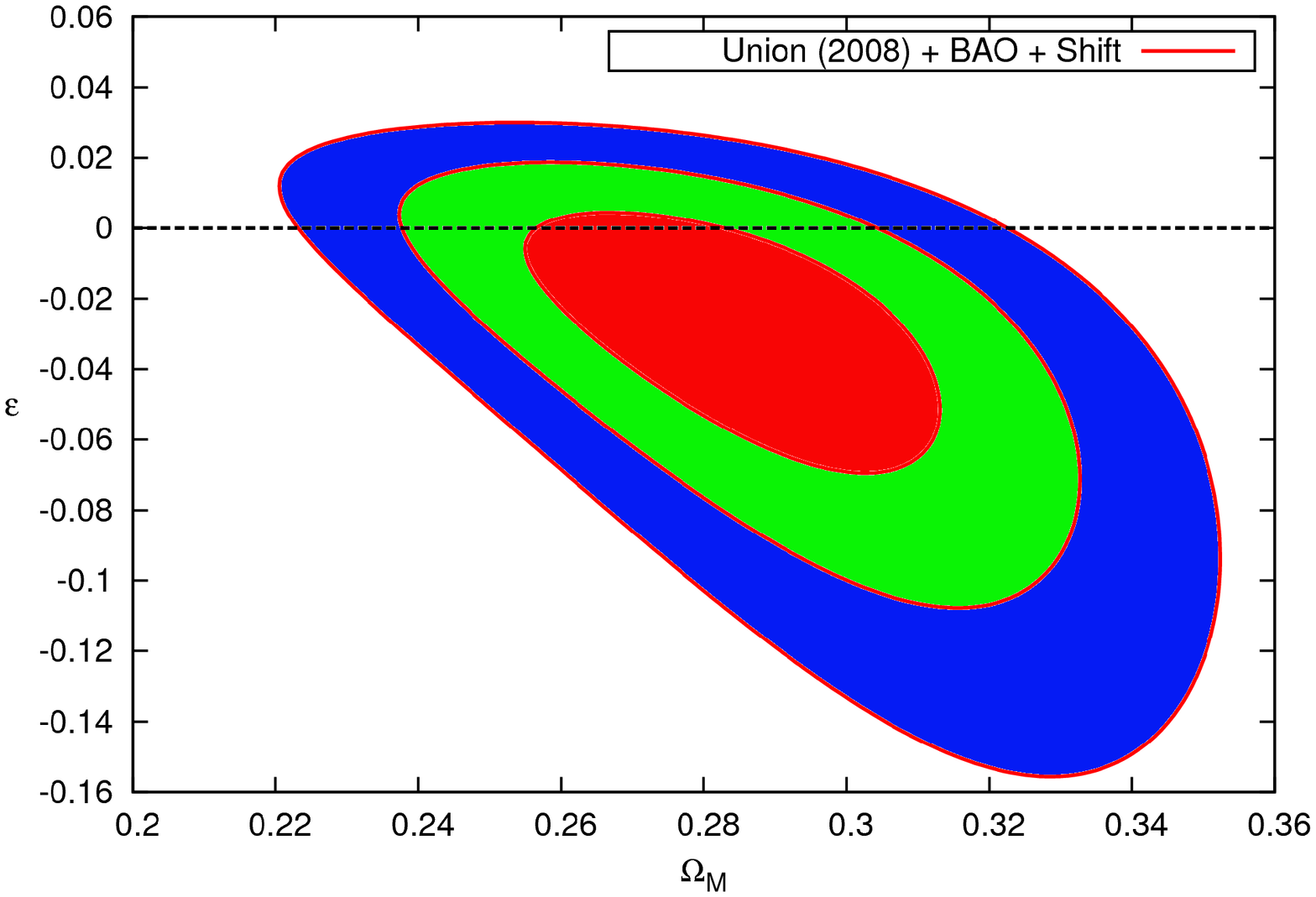,width=8cm,height=6.5cm}
\epsfig{file=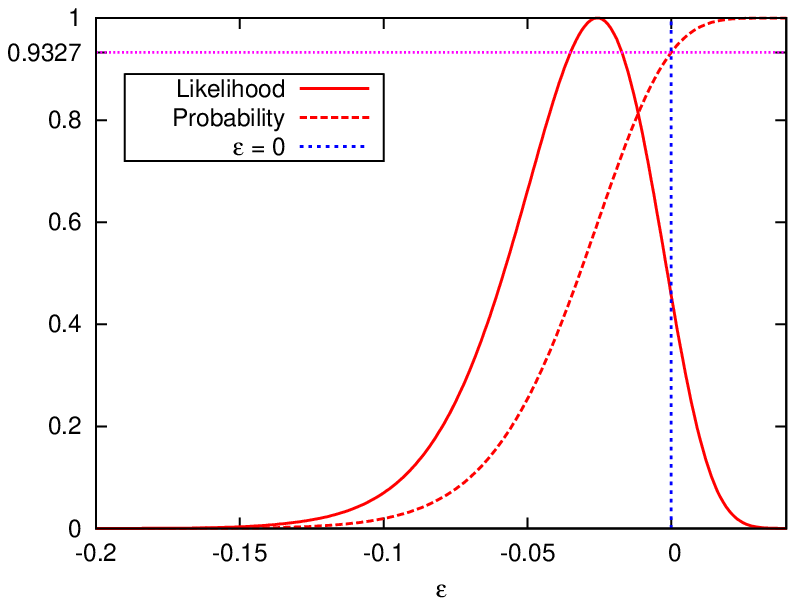,width=8cm,height=6.5cm}\hskip 0.1in}
 \caption{{\bf a)} Plane $\Omega_m$ - $\varepsilon$ showing the 68.3\%, 95.4\% and 99.7\% contours of probability, as well as the $\varepsilon=0$ line (dashed line). {\bf b)} Marginalized likelihood, normalized at the peak (solid line) and CDF (long-dashed line) for the $\varepsilon$ parameter, as well as the $\varepsilon=0$ line (short-dashed line) and $P(\varepsilon>0)$ line (dotted line). See text for details.}
 \label{fig:contlike}
\end{figure*}

The Baryon Acoustic Oscillations (BAO) given by the acoustic oscillations of baryons in the primordial plasma, leave a signature on the correlation function of galaxies as observed by Eisenstein {\it et al.} (2005) \cite{bao}. This signature furnishes a standard rule which can be used to constrain the following quantity \cite{bao}:
\begin{eqnarray}
{\cal{A}} \equiv \frac{\Omega_m^{1/2}}{
{{\cal{H}}(z_{\rm{*}})}^{1/3}}\left[\frac{1}{z_{\rm{*}}}
\Gamma(z_*)\right]^{2/3}  = 0.469 \pm 0.017, 
\label{A}
\end{eqnarray}
where ${\cal{H}}\equiv H(z)/H_0$, which is given by Eq. (\ref{friedmann}), $z_*=0.35$ is a typical redshift of the SDSS sample, and $\Gamma(z_*)$ is the dimensionless comoving distance to the redshift $z_*$. This quantity has been constrained by Eisenstein {\it et al.} \cite{bao} to be ${\cal{A}}=0.469\pm0.017$ and we shall use a gaussian prior on it.

SNe Ia luminosity distances constrain mainly the dark energy density and dark energy EoS, and we choose to use the up-to-date greatest combination of known SNs data sets, namely the Union compilation \cite{Union}. This sample is composed of 307 SNs, which includes the recent large samples of SNe Ia from the Supernova Legacy Survey \cite{ast05} and ESSENCE Survey \cite{essence}, older datasets, as well as the recently extended dataset of distant supernovae observed with Hubble Space Telescope (HST). We choose to use the ``without systematics'' Union dataset. We use a fairly well known \cite{Lewis02} analytical marginalization over $h$ to combine the SNe luminosity distances data.

An useful quantity to characterize the position of the Cosmic Microwave Background (CMB) power spectrum first peak is the shift parameter, which is given, for a flat Universe, by \cite{efstathiou}:
\begin{equation}
 {\cal R}=\sqrt{\Omega_m}\int_0^{z_r}\frac{dz}{{\cal H}(z)}
\end{equation}
where $z_r$ is the recombination redshift. The recombination redshift can be calculated by using the fitting formulas by Hu and Sugiyama \cite{hu}. The CMB shift parameter given by the WMAP 5-yr Monte Carlo Markov Chain (MCMC) analysis, assuming a standard (non-interacting) XCDM model is \cite{Komatsu08} ${\cal R}=1.710\pm0.019$.

As we marginalize over $h$ on the SNe data analysis, the only dependence on the Hubble constant comes from the shift parameter, via recombination redshift. But this dependence is too weak, so we choose to fix $h$ from the HST Key Project best fit, $h=0.72\pm0.08$ \cite{freedman}. So, finally, we have to minimize the following quantity:
\begin{equation}
\chi^2 = \tilde{\chi}^2_{SN} + \left(\frac{{\cal A}-0.469}{0.017}\right)^2+\left(\frac{{\cal R}-1.710}{0.019}\right)^2
\label{chi2}
\end{equation}
where $\tilde{\chi}^2_{SN}$ is the marginalized $\chi^2$ from the SN data, given by $\tilde{\chi}^2_{SN}=-2\ln\int_0^\infty{\cal L}_{SN}dh$, ${\cal L}_{SN}$ is the SNe magnitude likelihood, ${\cal L}_{SN}=\exp\left[-\frac{1}{2}\sum_{i=1}^{307}\left(\frac{\mu_{obs,i}-\mu_{th,i}}{\sigma_{\mu,i}}\right)^2\right]$, $\mu$ is the distance modulus, given as a function of the luminosity distance $d_L$ in Mpc as $\mu=5\log(d_L)+25$.

The total likelihood is given by ${\cal L}\propto e^{-\chi^2/2}$. Marginalizing ${\cal L}$ over $\omega$, we can see how much $\varepsilon$ and $\Omega_m$ can be constrained by the data. The 1, 2 and $3-\sigma$ contour levels are shown on Fig. \ref{fig:contlike}a.

We have found, then, from this analysis, $\Omega_m=0.284^{+0.019+0.039+0.059}_{-0.020-0.038-0.056}$ and $\varepsilon=-0.026^{+0.021+0.038+0.051}_{-0.027-0.061-0.106}$, for 1, 2 and 3$\sigma$, respectively. We have obtained a relative $\chi^2/\nu=311.66/305=1.02$, where $\nu$ is the number of degrees of freedom. As one can already see on Fig. \ref{fig:contlike}a, the possibility of $\varepsilon<0$ is clearly favored over the possibility of $\varepsilon>0$. As one may also see, the standard value, $\epsilon=0$, is marginally inside the 1-$\sigma$ region, in the joint analysis. In order to give a more quantitative result, we also marginalize over $\Omega_m$ with the aim to find the likelihood of $\varepsilon$, $\tilde{\cal L}(\varepsilon)$. This is shown on Fig. \ref{fig:contlike}b.

As one may see, the largest area under this curve is given for $\varepsilon<0$. It can be confirmed by calculating the probability $P(\varepsilon<0)$, which is given by
\begin{equation}
 P(\varepsilon<\varepsilon')=\int_{-\infty}^{\varepsilon'}\tilde{\cal L}(\varepsilon)d\varepsilon
\label{cdf}
\end{equation}

This is nothing more than the cumulative distribution function (CDF) of $\varepsilon$ and it is also shown on Fig. \ref{fig:contlike}b. By putting $\varepsilon'=0$ on (\ref{cdf}), we find that $P(\varepsilon<0)=93.27\%$. It corresponds to a ratio $P(\varepsilon<0)/P(\varepsilon>0)=13.86$, and an almost 2-$\sigma$ c.l. in favor of $\varepsilon<0$ in comparison with $\varepsilon>0$. Here, on the marginalized analysis, we can also see that the standard value, $\epsilon=0$, is outside the 1-$\sigma$ region, but is well inside the 2-$\sigma$ region, though.

\section{Conclusions}
Assuming that
Dark Energy is amenable to a fluid description with a well defined
temperature and chemical potential, we have analyzed the possibility of having DM decaying into DE through thermodynamical considerations. We have shown that, once at least one of the fluids is provided with a non null chemical potential, this possibility is allowed by the second law of thermodynamics. We have also showed that the Le Ch\^atelier-Braun principle remains valid. Next we have applied this result to put observational constraints on a fairly simple model of interaction between DM and DE. We have found that, in the context of this model, the decay of DM into DE is reasonably favored over the inverse process. This model has the correct behavior of being DM dominated in the past and DE dominated in the future. Our result gives a theoretical basis and opens the possibility of extending and analyzing other models of interacting DE present on the literature. However, it should be mentioned that our results rely on the assumption of a nearly standard evolution of pertubations on interacting DE models, as indicated by \cite{Abdalla08}. While this result applies to the model (\ref{Q}), a more general perturbation analysis is still lacking. Work in this direction is being done and can appear in a forthcomming communication.

\section*{Acknowledgements}
SHP is supported by CNPq No. 150920/2007-5 (Brazilian Research Agency). JFJ is supported by CNPq. JFJ would like to thank Prof. L. R. W. Abramo for useful comments about interacting DE. We would like to thank Prof. J. A. S. Lima and R. L. Holanda for useful comments.

\end{document}